\documentclass[runningheads]{llncs}
\usepackage{graphicx}

\usepackage{caption}
\usepackage{subcaption}
\usepackage{easyReview}

\begin{document}
\title{Simulating Delay in Seeking Treatment for Stroke due to COVID-19 Concerns with a Hybrid Agent-Based and Equation-Based Model}
\titlerunning{Delay in Stroke Treatment due to COVID-19}
%
\author{Elizabeth Hunter\inst{1}\orcidID{0000-0002-1767-4744} \and
Bryony L. McGarry\inst{1,2}\orcidID{0000-0002-8784-2109} \and
John D. Kelleher\inst{1,3}\orcidID{0000-0001-6462-3248}}
\authorrunning{E.Hunter et al.}
%
\institute{PRECISE4Q Predictive Modelling in Stroke, Technological University Dublin, Dublin, Ireland  
\and
School of Psychological Science, University of Bristol, Bristol, United Kingdom
%
\and 
ADAPT Research Centre, Technological University Dublin, Dublin, Ireland \\
\email{elizabeth.hunter@tudublin.ie}}

\maketitle              
\begin{abstract}

COVID-19 has caused tremendous strain on healthcare systems worldwide. At the same time, concern within the population over this strain and the chances of becoming infected has potentially reduced the likelihood of people seeking medical treatment for other health events. Stroke is a medical emergency and swift treatment can make a large difference in patient outcomes.  Understanding how concern over the COVID-19 pandemic might impact the time delay in seeking treatment after a stroke can be important in understanding both the long term cost implications and how to target individuals during another pandemic scenario to remind them of the importance of seeking treatment immediately. We present a hybrid agent-based and equation-based model to simulate the delay in seeking treatment for stroke due to concerns over COVID-19 and show that even small changes in behaviour impact the average delay in seeking treatment for the population. This delay could potentially impact the outcomes for stroke patients and future healthcare costs to support them. We find that introducing control measures and having multiple smaller peaks of the pandemic results in less delay in seeking treatment compared to a scenario with one large peak. 

\keywords{Agent-based model \and hybrid model \and Stroke \and COVID-19.}
\end{abstract}
\section*{}
This is a preprint of the following chapter: Hunter, E. McGarry, B.L, and Kelleher, J.D., Simulating Delay in Seeking Treatment for Stroke Due to COVID-19 Concerns with a Hybrid Agent-Based and Equation-Based Model, published in Advance in Social Simulation: Proceedings of the 16th Social Simulation Conference, 20–24 September 2021, edited by  Marcin Czupryna, Bogumił Kamiński, 2022, Springer, reproduced with permission of Springer Nature Switzerland AG 2022. The final authenticated version is available online at: 10.1007/978-3-030-92843-8$\_$29

\section{Introduction}

 The COVID-19 pandemic has a wider impact than just those infected by the virus.  All parts of society have been affected, and it has been shown in many countries that there are additional excess deaths during the pandemic that are not directly explained by the deaths due to COVID-19 infection~\cite{Brant1898,Woolf2020}.  Thus, the full impact of the pandemic on the health care system is yet to be determined.  There is speculation that excess deaths may be attributable to a delay in seeking treatment of many non-communicable diseases due to concerns over hospital overcrowding and being exposed to the virus which lead to a `watch-and-wait' approach  ~\cite{liu_global_2020,hoyer_acute_2020}.  Here we focus on the impact of COVID-19 on treatment seeking behaviour of stroke sufferers, as it is one disease where minimising the time from symptom onset to treatment is particularly crucial to the patient's outcome~\cite{matsuo_association_2017}.   

Worldwide, stroke  is the second leading cause of death, a major cause of disability~\cite{johnson_global_2019} and one of the most expensive neurological conditions~\cite{rajsic_economic_2019}. Ischaemic strokes are the most common~\cite{johnson_global_2019} and are defined as an episode of neurological dysfunction caused by focal cerebral infarction~\cite{sacco_updated_2013}. Early treatment has been associated with better outcomes~\cite{lees_time_2010,gumbinger_time_2014}. Stroke caused by intracerebral haemorrhage (ICH) is an episode of neurological dysfunction caused by a non-traumatic bleed within the brain or ventricular system~\cite{sacco_updated_2013}. Rapid care has been associated with improved case-fatality rates~\cite{mcgurgan_acute_2021,parry-jones2019}. Both ischemic and stroke caused by ICH are considered neurological emergencies and require fast diagnosis and treatment to minimise the short and long-term health impacts~\cite{mcgurgan_acute_2021}. 

Estimating the time delay in seeking treatment across stroke patients caused by concerns about COVID-19 highlights the cost of COVID-19 on stroke treatment, including COVID-19's effect on direct costs such as acute stroke care and long-term hospitalisation and treatment (medication, physiotherapy), rehabilitation, reintegration and quality of life and indirect costs such as productivity loss from unemployment, informal caregiving and premature mortality~\cite{girotra_contemporary_2020,joo_literature_2014}.  

In this paper, we use a hybrid agent-based and equation-based model to simulate the delays in seeking medical care for a stroke due to concern about the COVID-19 pandemic. In the following sections, we describe the model, then discuss the experiments run with the model and finally, the results.

\section{Model}

At its core, the model is an agent-based model that simulates stroke incidence within a population   to determine the delay in seeking stroke treatment. Agent-based models for infectious disease spread have four main components, Society, Environment, Disease and Transportation~\cite{hunter2017}. Although stroke is not an infectious disease our model also has those four main components. The next sections describe the components in more detail.  

\subsection{Society}

The society component uses Irish census data~\cite{CSO_sa} to accurately simulate the age and sex distribution of those 50 and older in Ireland.\footnote{Those age 50 and older make up approximately 30.4$\%$ of the Irish population.} We choose those 50 and older as the majority of strokes in Ireland occur in those over the age of 65~\cite{Strokeregister} and models predicting stroke and cardiovascular risk often start at the ages of 40 or later~\cite{conroy_estimation_2003,dagostino_stroke_1994}. Each agent has both an age and a sex. Table \ref{table:agents} shows the number of agents in three age groups (under 65, 65 to 79, and over 80) by sex. 

\begin{table}
 \centering
\caption{Age and Sex breakdown of Agents in the model.}\label{table:agents}
\begin{tabular}{|l|l|l|}
\hline Age Group & Male & Female \\
\hline
Under 65& 400,768 & 408,125 \\
65 to 79& 238,579 & 250,396 \\
Over 80 & 58,258 & 90,334\\
\hline
\end{tabular}
\end{table}

\subsection{Environment}
The environment component of our model is an equation-based component. Typically the environment component of an agent-based model for infectious disease spread represents a geography that agents move through. However, instead of representing the geography of a region, here we consider the environment to be cases of COVID-19. We do not consider the COVID-19 disease status of the agents in the model, but each agent knows the number of people who have recently tested positive for COVID-19 in the country. As the number of cases increases so to does the agent's concern regarding the pandemic and their hesitancy in seeking medical treatment post stroke. 

Concern is discussed more in Section \ref{sec:transportation}. 
To determine the number of COVID-19 cases, we use a difference equation model based on the Irish SEIR population level model~\cite{Gleeson2020}. We use difference equations instead of differential equations because the difference equations use discrete time steps and are thus more analogous to the agent-based model. 

 Each time step in the model is a day and for each time step the difference equations determines the number of agents who have tested positive. This number informs the agents of the current state of the pandemic in Ireland and factors into their decision on waiting to go to the hospital after having a stroke.

Although the agents only represent the population in Ireland over age 50, the COVID-19 SEIR model is run for all of Ireland. The initial conditions are set to roughly mimic the start of the pandemic when there are only a few agents infected in the country. The model starts with 4,937,769 susceptible agents; 6 exposed agents; 12 infectious agents and 0 recovered agents. Of the 12 infectious, 1 is pre-symptomatic, 5 are asymptomatic, 1 is isolating, 2 are waiting for tests and isolating, 2 have tested positive and 1 is not isolating.

\subsection{Disease}
The disease component of the model determines the stroke incidence in the population.  This component determines if an agent will have a stroke or not on a day within the model and has been created so that the incidence of stroke in the agent population matches stroke incidence in the actual population. We use an estimate of absolute stroke risk for agents. This risk is calculated for the six different groups of agents outlined in Table \ref{table:agents}.  

Absolute risk is defined as the number of events, in this case, strokes, in the group divided by the total number of people in the group. To get the total number of people in each of our six groups, we use Irish Census data, and to get the number of strokes in each group, we use statistics from the National Stroke Register Report 2018~\cite{Strokeregister}, which provides the total number of strokes in Ireland in 2018 that were males or females and then the percent of strokes in the three age groups (less than 65, 65 to 79 and over 80). 

Table \ref{table:absolute risk} shows the absolute risk for the six age and sex categories within a year. To determine the risk of stroke on a given day, we divide the yearly risk by 365. Then at each time step, each agent samples a number from a uniform probability distribution between 0 and 1. If the number sampled is less than the daily risk of stroke for their demographic, the agent will have a stroke.

\begin{table}
\centering
\caption{Absolute Risk of Stroke by Age and Sex.}\label{table:absolute risk}
\begin{tabular}{|l|l|l|}
\hline
Age Group & Male & Female \\
\hline
Under 65& 0.1$\%$  & 0.05$\%$ \\
65 to 79& 0.4$\%$ & 0.2$\%$ \\
Over 80 & 0.9$\%$ & 0.7$\%$\\
\hline
\end{tabular}
\end{table}

\subsection{Transportation} 
\label{sec:transportation}
The transportation component of the model determines the delay in the agents who have had a stroke arriving at the hospital for treatment.  
After a stroke, an agent chooses on of four behaviours (Self Treat, Wait and See, Seek Advice, Seek Medical Advice) and each behaviour results in a different average time to hospital arrival. The behaviours, the resulting time delays in seeking treatment, and the percent of strokes that follow each behaviour are determined from~\cite{mandelzweig_lori_perceptual_2006}. Table \ref{table:behaviours} shows the baseline percent chance that an agent who had a stroke will follow each of the four behaviours and the resulting time to reach the hospital.  After a stroke, an agent decides which behaviour they will take and this determines their personal delay to treatment time. 

To make this decision an agent samples a number from a uniform probability distribution between 0 and 1. They then use thresholds based on the percents in Table \ref{table:behaviours} to determine which behaviour they will choose.  If the random-number is less than  0.06 they will self-treat; if it is greater than 0.06 and less than 0.38 they will wait and see; if it is greater than 0.38 and less than 0.50 they will seek advice and if it is greater than 0.50 they will seek medical advice. The probabilities that an agent will choose each of the behaviours presented here are for the baseline scenario and will change based on the level of concern an agent has over the number of COVID-19 cases in the environment.   

\begin{table}
\centering
\caption{Post Stroke Behaviours and Time to Reach the Hospital~\cite{mandelzweig_lori_perceptual_2006}  }\label{table:behaviours}
\begin{tabular}{|l|l|l|}
\hline
Behaviour & Percent & Time \\
\hline
Self Treat & 6 & 11.5  \\
Wait and See &32 & 13.25 \\
Seek Advice & 12 & 4.25\\
Seek Medical Advice & 50 &  2\\
\hline
\end{tabular}
\end{table}

As the number of cases of COVID-19 increase, agents become more concerned about the pandemic. The concern levels of agents are based on the \emph{Am\'arach} public health surveys in Ireland~\cite{amarach}. Concern levels from the survey show higher levels of worry occurring around the peaks of the different waves of the pandemic. 
Table \ref{table:concern} list the thresholds for changing the mean concern level in the model.

\begin{table}
\centering
\caption{Thresholds for different level of agent Concern.}\label{table:concern}
\begin{tabular}{|l|l|l|l|l|}
\hline
Number of Cases & Concern & Number of Cases & Concern\\
\hline
Less than 50 & 0 & 3000 to 10000 & 7.2 \\
50 to 100 &  5 &10000 to 15000 & 7.3\\
100 to 200 & 5.6 &15000 to 20000 & 7.4\\
200 to 1000 & 6 & 20000 to 21000 & 7.5\\
1000 to 3000 & 7 &21000 plus & 8\\

\hline
\end{tabular}
\end{table}

We do not assume that all agents have the same level of concern. Agents are assigned a concern level using a normal distribution with the mean concern level from Table \ref{table:concern} and a standard deviation of 0.5. The higher the concern level of the agent, the less likely they are to seek medical advice and the more likely they are to self-treat, wait and see or seek non-medical advice. Agents behaviours and concern are not related to demographics with all agents equally likely to select a given behaviour or concern. 

\subsection{Schedule}
The model runs on discrete time steps with each time step equating to a single day in the model. At each time step the COVID-19 model determines the number of people who have tested positive in the country, the agents determine their level of concern, if they had a stroke on that day and how long the delay was before they reached the hospital. 
\section{Experiment}

To look at how the level of agent concern impacts the average delay time in seeking treatment after a stroke we run a number of different scenarios.
\begin{enumerate}
    \item A baseline scenario with no COVID-19 cases.
    \item A scenario with no COVID-19 restrictions and there is one peak of cases.
    \item A scenario with rolling lockdowns that results in multiple peaks of cases.
\end{enumerate}


Scenarios 2 and 3 are run twice to account for different levels of behaviour change based on concern. One a low level of behaviour change where agents are only slightly less likely to seek medical advice, and one high-level change where agents are much less likely to seek medical advice. In all three scenarios agents choose their behavioural response to a stroke using the method discussed in the transportation section. What varies between the scenarios is the probability of each behaviour. Table \ref{table:concernbehaviour} shows the probability an agent will choose a behaviour based scenario and their concern.
In this table, each cell in the behaviour columns records the probability of the behaviour, and (in brackets) the thresholds defining the interval that the random number an agent samples must fall within in order the agent to adopt that behaviour. 
For each scenario, the model is run 25 times to account for stochasticity in the model. The number of runs was determined using the method in~\cite{Hunter2020}.

\begin{table}
\centering
\caption{Probability (and sample interval) for behaviour by scenario and concern}\label{table:concernbehaviour}
\begin{tabular}{|l|l|l|l|l|l|}
\hline
Scenario & Concern & Self Care & Wait $\&$ See & Non-Medical Advice & Medical Advice \\
\hline

\textbf{Low level of}&less than 5 & 0.06 \textit{(0, 0.06)} & 0.32 \textit{(0.06, 0.38)} & 0.12 \textit{(0.38, 0.50)}& 0.50 \textit{(0.50, 1)}\\
\textbf{Behavioural}&5 &0.07 \textit{(0, 0.07)}& 0.33 \textit{(0.07, 0.40)}& 0.13 \textit{(0.40, 0.53)}& 0.47 \textit{(0.53, 1)}\\
\textbf{Change}&6 & 0.08 \textit{(0, 0.08)}& 0.34 \textit{(0.08, 0.42)}& 0.13 \textit{(0.42, 0.55)} & 0.45 \textit{(0.55, 1)}\\
\textbf{with Concern}&7 and greater & 0.09 \textit{(0, 0.09)}& 0.35 \textit{(0.09, 0.44)}& 0.15 \textit{(0.44, 0.59)}& 0.41 \textit{(0.59, 1)}\\
\hline
\textbf{High level of}&less than 5 & 0.06 \textit{(0, 0.06)}& 0.32 \textit{(0.06, 0.38)}& 0.12 \textit{(0.38, 0.50)} & 0.50 \textit{(0.50, 1)}\\
\textbf{Behavioural}&5 & 0.08 \textit{(0, 0.08)}&0.34 \textit{(0.08, 0.42)}& 0.14 \textit{(0.42, 0.56)}& 0.44 \textit{(0.56, 1)}\\
\textbf{Change}&6 & 0.10 \textit{(0, 0.10)}& 0.36 \textit{(0.10, 0.46)}& 0.16 \textit{(0.46, 0.62)}& 0.38 \textit{(0.62, 1)}\\
\textbf{with Concern}&7 and greater & 0.12 \textit{(0, 0.12)}& 0.38 \textit{(0.12, 0.50)}& 0.18 \textit{(0.5, 0.68)}& 0.32 \textit{(0.68, 1)} \\
\hline
\end{tabular}
\end{table}


\section{Results}
The following sections discuss the results from the different experiments. For each scenario, we look at the average delay in seeking treatment for stroke over the year. The average delay is examined in relation to the number of COVID-19 cases and the average concern across all agents in the model for a given day.  Finally, we compare all of the scenarios to look at the potential increase in delay in treatment due to concern about the COVID-19 pandemic. 

\subsection{Baseline Scenario}

The baseline scenario for our model is a situation where there are no background cases of COVID-19 as such, the agents do not have any concern about hospital overcrowding or COVID-19 infection at the hospital.  Thus agents follow the post-stroke behaviours and timing in Table \ref{table:behaviours}. To look at the delay in seeking stroke treatment, we first look at the average delay across a whole model run or a whole year.  For each run, we find the average delay across all agents who have had a stroke during the year and then take the average of that across the 25 model runs.  Similarly, we find the median delay for all agents who have had a stroke during the year and then take the average of the medians across the 25 model runs. Table \ref{table:baselineavgmedian} shows the average median delay for seeking treatment and the average delay for seeking treatment across the 25 runs, as well as the maximum and minimum values and standard deviation.

\begin{table}
\centering
\caption{Delay in Seeking Stroke Treatment in the Baseline Scenario.}\label{table:baselineavgmedian}
\begin{tabular}{|l|l|l|}
\hline
 & Average Delay (hrs) & Median Delay (hrs)\\
\hline
Average & 6.4 &  2.9 \\
Maximum & 6.7 & 2.0 \\
Minimum & 6.2 & 4.3\\
Standard Deviation & 0.14 & 0.80\\

\hline
\end{tabular}
\end{table}

From the table, we can see that while there is a higher average delay, the median is lower, showing that 50$\%$ of agents have stroke treatment within 2.9 hours.  This makes sense based on the agents’ responses to stroke determined from Table \ref{table:behaviours} where 50$\%$ of agents should seek medical care upon having a stroke which would result in a delay of approximately 2 hours to treatment.

\subsection{Single Peak}

\begin{figure}
\begin{subfigure}{.5\textwidth}
  \centering
  \includegraphics[width=\linewidth]{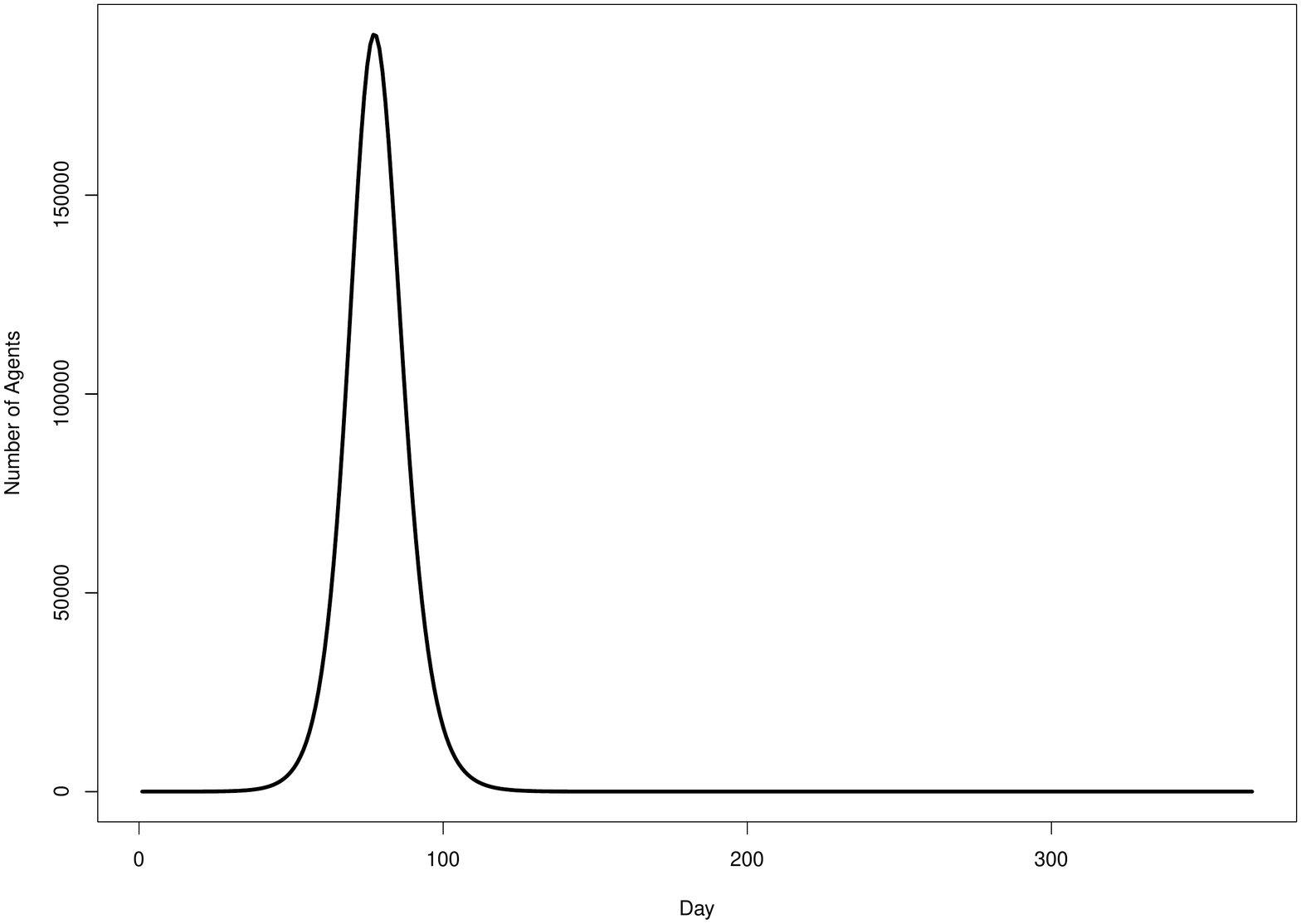}
  \caption{Total  tested positive per day.} \label{fig:Peakcases}
\end{subfigure}%
\begin{subfigure}{.5\textwidth}
  \centering
  \includegraphics[width=\linewidth]{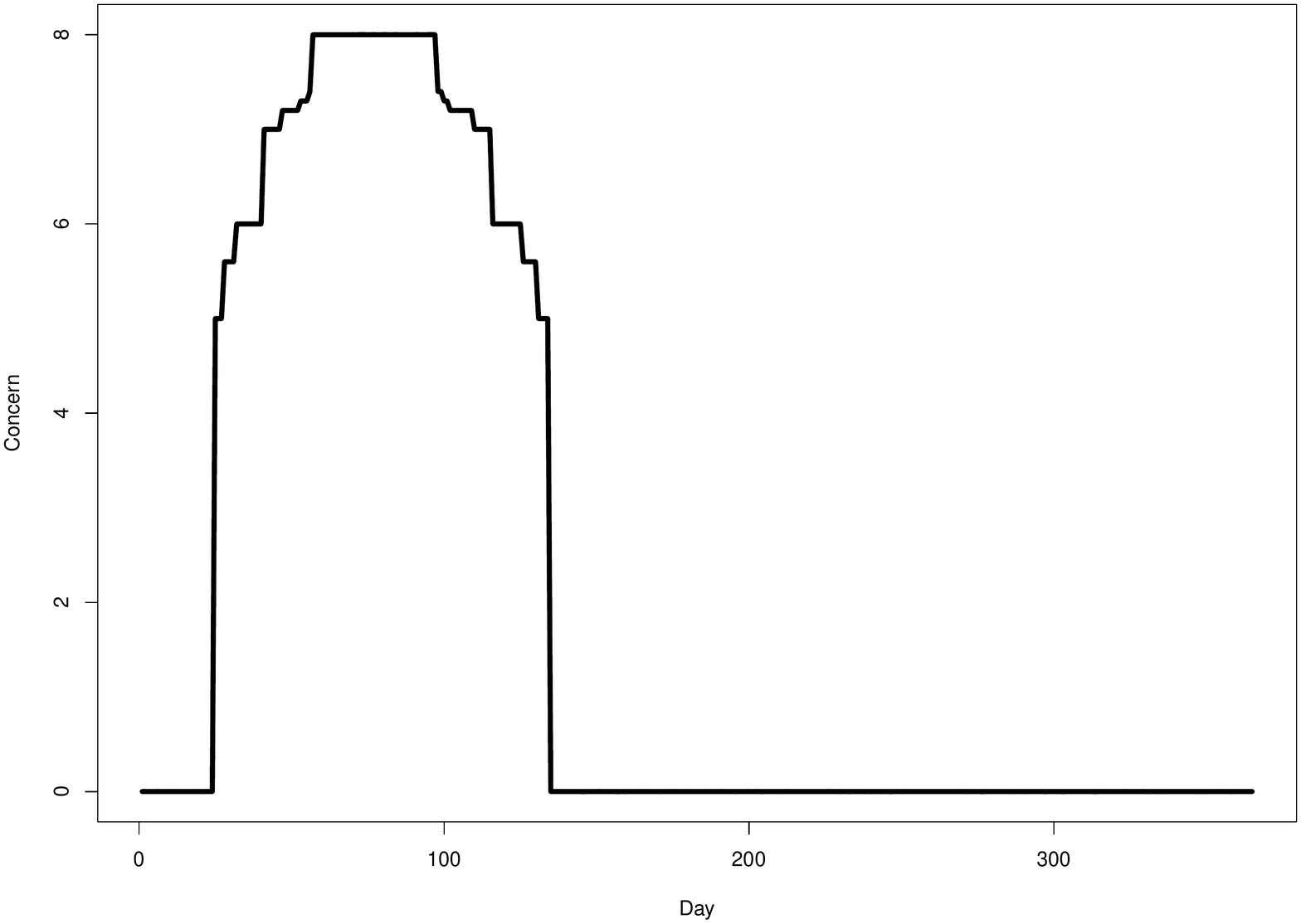}
\caption{Average level of concern.} \label{fig:Peakconcern}
\end{subfigure}
\caption{Total cases tested positive by day and average concern by day.}
\label{fig:test}
\end{figure}

The next scenario we look at is a scenario that has a single high peak of COVID-19 cases. This scenario would correspond to a real world situation where no intervention measures were taken to slow the spread of COVID-19.  Figure \ref{fig:Peakcases} shows the total number of cases tested positive for COVID-19 in this scenario, and Figure \ref{fig:Peakconcern} shows the average level of concern across all agents due to COVID-19 in the scenario.  

 The high number of cases and high concern in the first 100 days of the year results in agents adapting their behaviours upon having a stroke.  To account for different levels of behaviour change, we look at both a high and low behaviour change scenario. The corresponding changes in behaviour for high and low impact were previously discussed in Table \ref{table:concernbehaviour}. Table \ref{table:peakavgmedian} shows the average delay in seeking treatment and the median delay in seeking treatment across the 25 runs for both the high and low impact scenarios. From the table, we can see that there is a slight increase in the average delay across the runs and average of the median delay across the runs going from a scenario where concern over COVID-19 has a low impact on behaviours post-stroke versus a higher impact.   

\begin{table}
\centering
\caption{Delay in Seeking Stroke Treatment with a Single Peak of COVID-19.}\label{table:peakavgmedian}
\begin{tabular}{|l|l|l|l|l|l|}
\hline
& Behaviour Change& Average & Maximum & Minimum & Standard Deviation \\

\hline
\textbf{Average Delay} & High & 6.9 & 6.6 & 7.3 & 0.14 \\
&Low  & 6.7 & 6.5 & 6.9 & 0.10 \\
\hline
\textbf{Median Delay} & High  & 4.2 & 4.0 & 4.5 & 0.12\\
&Low  & 4.1 & 3.6 & 4.5 & 0.18 \\

\hline
\end{tabular}
\end{table}
\subsection{Multiple Peaks}

\begin{figure}
\begin{subfigure}{.5\textwidth}
  \centering
  \includegraphics[width=\linewidth]{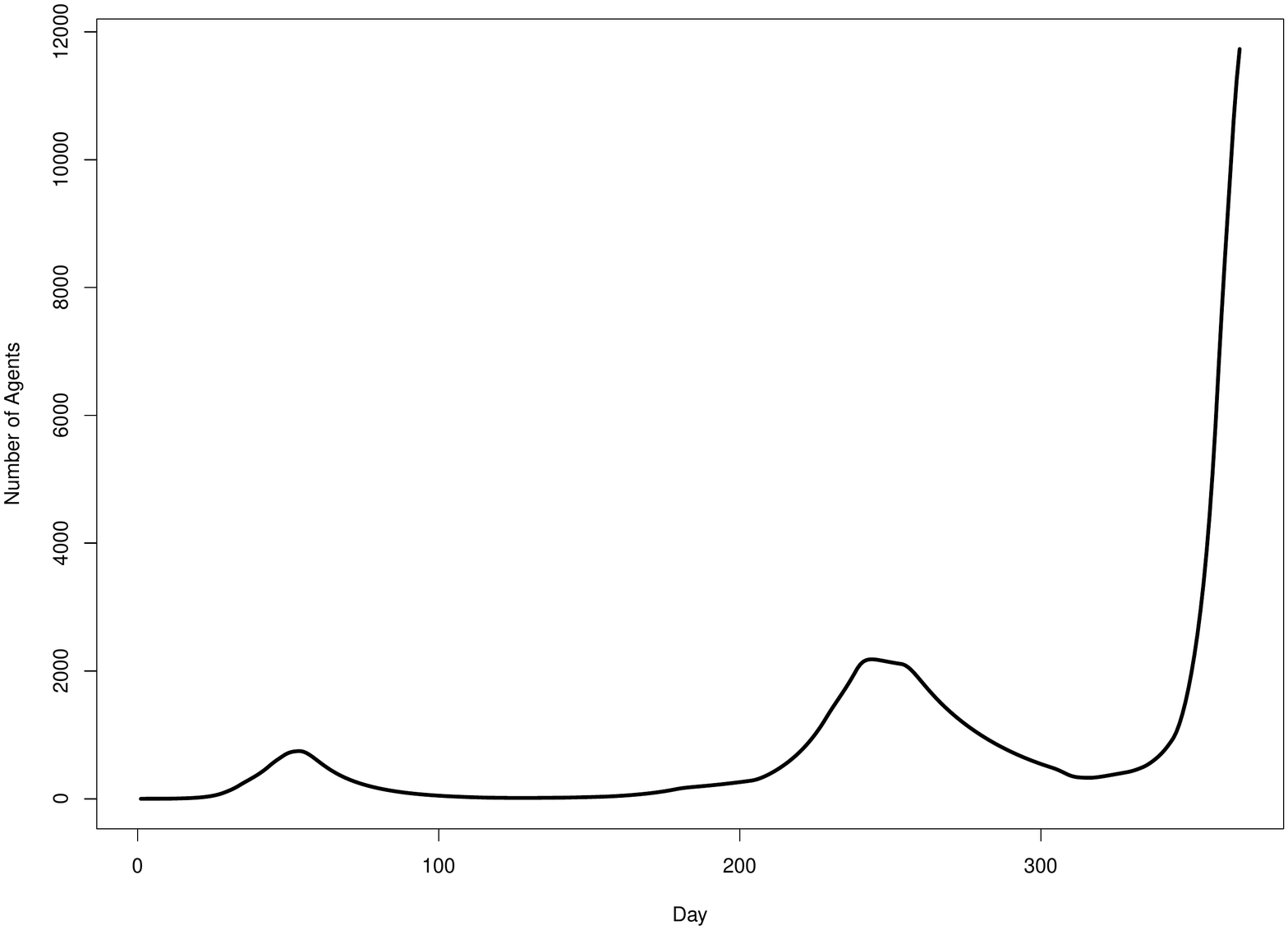}
  \caption{Total tested positive per day.} \label{fig:Rollingcases}
\end{subfigure}%
\begin{subfigure}{.5\textwidth}
  \centering
  \includegraphics[width=\linewidth]{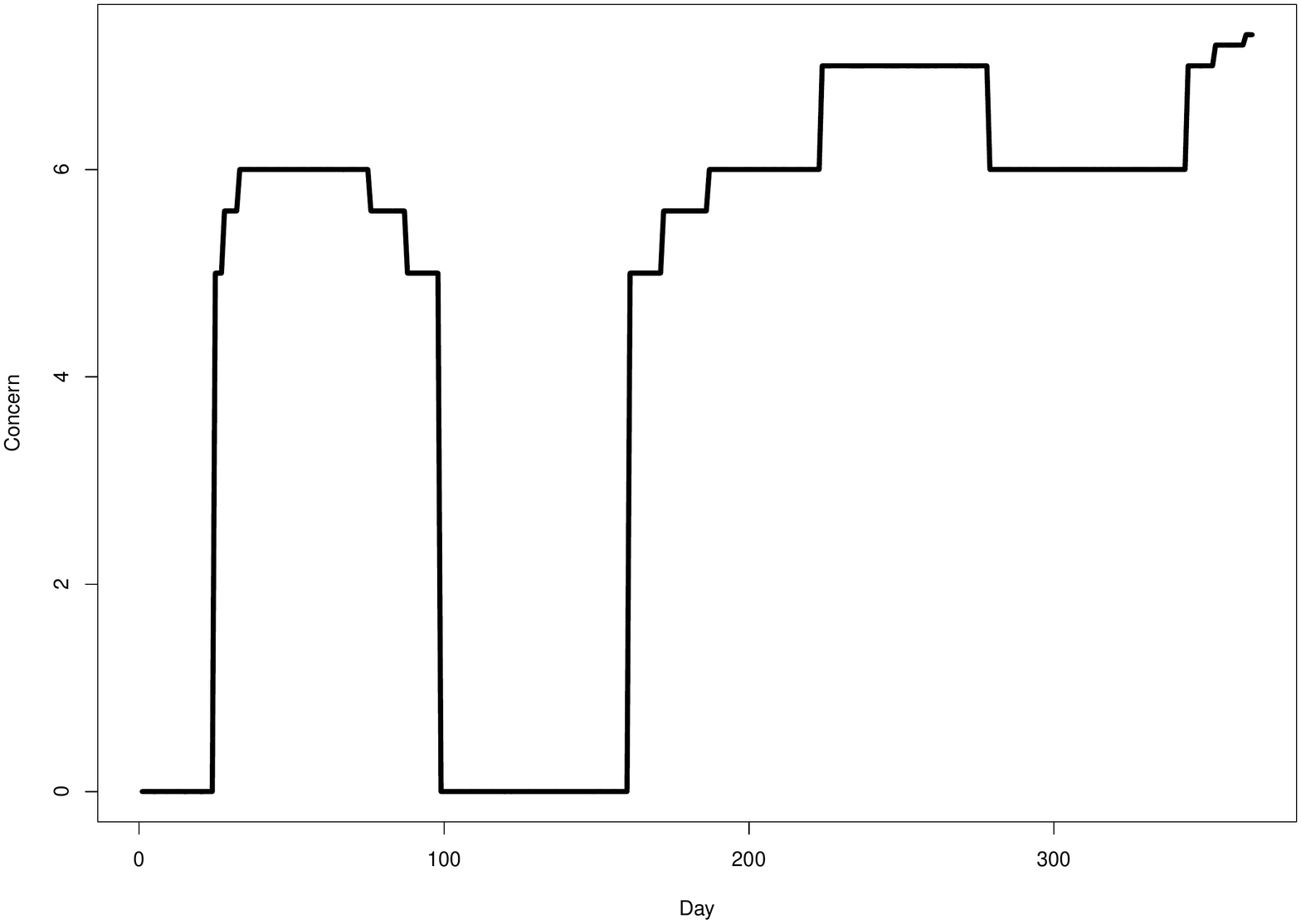}
\caption{Average concern.} \label{fig:Rollingconcern}
\end{subfigure}
\caption{Total cases tested positive by day and average concern by day.}
\label{fig:rolling}
\end{figure}

The next scenario that is more realistic in terms of COVID-19 cases represents a set of rolling restrictions that lead to multiple peaks in cases of COVID-19 throughout the year. The timing and size of the peaks were set to mimic the cases that occurred in Ireland during 2020.   Figure \ref{fig:Rollingcases} shows the total number of cases tested positive for COVID-19 in this scenario, and Figure \ref{fig:Rollingconcern} shows the average level of concern across all agents due to COVID-19 in the scenario.  With multiple infection peaks, we see a sustained high level of concern for most of the year.  However, while the concern levels in the scenario with one peak reach a maximum of about 8 in Figure \ref{fig:Peakconcern} the concern in the scenario with multiple peaks has a maximum of just under 7.  

Similar to the scenario with a single peak, we look at both a high and low behaviour change scenario.  Table \ref{table:rollingavgmedian} shows the average delay in seeking treatment and the median delay in seeking treatment across the 25 runs for both the high and low impact scenarios. From the table, we can see that the high and low impact on behaviours seems to have a larger impact on the average delay in seeking stroke treatment, whereas the median delay is not as impacted.

\begin{table}
\centering
\caption{Delay in Seeking Stroke Treatment with Multiple Peaks of COVID-19.}\label{table:rollingavgmedian}
\begin{tabular}{|l|l|l|l|l|l|}
\hline
&Behaviour Change & Average & Maximum & Minimum & Standard Deviation\\

\hline
\textbf{Average Delay} &High  &  6.8 & 6.6 & 7.3 & 0.17\\
&Low  &  6.6 & 6.4 & 6.9 &  0.11\\
\hline

\textbf{Median Delay} &High Impact & 4.2 & 4.0 & 4.5 & 0.14\\
& Low Impact &  4.2 & 3.8 & 4.4 & 0.15 \\

\hline
\end{tabular}
\end{table}

\subsection{Comparison of Scenarios} 

To compare the scenarios, we look at the percent change in the delay of seeking stroke treatment between the baseline and each of the other scenarios and do a one-sided t-test to determine if the average delay in treatment is greater in the COVID-19 scenarios compared to the baseline.  Table \ref{table:comparison} shows the percent change and p-value for each of the four scenarios discussed in the previous sections.  

\begin{table}
\centering
\caption{Change in Delay Between the Baseline and COVID-19 Scenarios.}\label{table:comparison}
\begin{tabular}{|l|l|l|l|}
\hline
& Behaviour Change & Percent Change  &  p-value  \\
\hline
\textbf{Single Peak} &High  & 7.8 &  2.2e-16 \\
&Low  & 4.7 & 1.6e-10\\
\hline
\textbf{Multiple Peaks} &High  & 6.3 & 1.9e-13\\
&Low &  3.1 & 5.0e-8\\

\hline
\end{tabular}
\end{table}
 
Looking at the table, we can see that for all of the t-tests comparing the average delay to the baseline, we have p-values that are much less than 0, showing that the average delays in the scenarios where agents change their behaviour due to COVID-19 are significantly greater than the average delay in the baseline scenario with no COVID-19 cases. Additionally, we see that even though the single peak scenario does not have sustained concern throughout the year as the multiple peaks scenario does, the higher level of concern during the single peak leads to a higher percent change in delay from the baseline. As expected for both the single peak and the multiple peaks scenarios, the versions where concern has a higher impact on agents behaviour results in a larger difference from the baseline compared to when concern has a lower impact on behaviour.

\section{Conclusion}

Our model results show that if stroke patients change their behaviours in seeking treatment after stroke symptoms due to concern over hospital capacity or the possibility of COVID-19 infection, there could be a significant increase in the delay in patients arriving at the hospital and thus a delay in receiving the appropriate treatment.  This is not to be confused with delay in acute care having arrived at the hospital, which may vary according to the hospital site. However, from a multi-centre study~\cite{hoyer_acute_2020}, time to treatment during the pandemic in acute stroke care did not differ from time to treat prior to the pandemic from pre-pandemic timing.  

The results show that even though the single peak scenario results in no cases and low concern after about 100 days, the higher levels of concern due to higher cases at the peak results in more delay in   seeking treatment.  This suggests that introducing measures to control the pandemic will not only save lives lost to COVID-19 but might also save lives lost to stroke.  

It is important to note that the model is not intended to predict the actual hours that individuals who have had a stroke are delayed in seeking treatment but to show the possible impact on delays. We see that even in scenarios with low behaviour change there is still an impact on the delay in seeking treatment. As it is essential to treat stroke as rapidly as possible to ensure the best outcomes, even a small delay in time when seeking treatment could have a significant influence on outcomes which may result in higher post-stroke medical costs. Our results highlight the importance of considering not just the impact of COVID-19 cases on the healthcare system but also on the long term impacts of the pandemic on stroke outcomes. In future scenarios where there are concerns about hospital capacity it may be necessary to target individuals at risk of stroke with information about the importance of seeking rapid medical care.   

Our model does not take into account any factors beyond a patient's age, gender, and the cases of COVID-19 in the background. The model could be improved and made more realistic by including additional factors.  For example, distance to a hospital could impact the delay in seeking treatment, and other risk factors, such as smoking or diabetes could impact an agent's risk of stroke.  Additionally, concern could change based on an agent's characteristics: the more likely an agent is to have severe COVID-19 the more concerned about the pandemic they will be. Delays in seeking treatment may also result from social distancing measures introduced during the COVID pandemic~\cite{hoyer_acute_2020}. In pre-pandemic times 96$\%$ of stroke emergency calls were activated by caregivers and witnesses~\cite{dhand_acute_2020,wein_activation_2000}. Instructions to avoid contact with others means there is less opportunity for the symptoms to be witnessed or disclosed to others~\cite{hoyer_acute_2020}.   

Before the COVID-19 pandemic, stroke sufferers  delays in seeking treatment resulted in worse functional outcomes of stroke patients~\cite{matsuo_association_2017}. This may be especially so during a pandemic, as patients with severe stroke would require longer hospitalisation, potentially increasing their exposure to in-hospital pathogens and placing further constraints on hospital resources~\cite{khosravani_protected_2020}. 

\textbf{Funding:} This project received funding from the EU’s Horizon 2020 research and innovation programme under grant agreement No. 777107, and by the ADAPT Centre for Digital Content Technology funded under the SFI Research Centres Programme (Grant 13/RC/2106) and co-funded under the European Regional Development Funds.

%
%

%
%
%
 \bibliographystyle{splncs04}
\bibliography{strokedelay.bib}

\end{document}